# Correct Small-Truncated Excited State Wave functions Obtained via Minimization Principle for Excited States compared / opposed to Hylleraas-Undheim and McDonald higher "roots"

Z. Xiong[1,2, b], J. Zang[1], H.J. Liu[1], D. Karaoulanis[3], Q. Zhou[2] and N.C. Bacalis[4, a]

[1]*Space Science and Technology Research Institute, Southeast University, Nanjing, 21006, Peoples Republic of China*
[2]*School of Economics and Management, Southeast University, Nanjing 210096, Peoples Republic of China*
[3]*Korai 21, Halandri, GR-15233, Greece.*
[4]*Theoretical and Physical Chemistry Institute, National Hellenic Research Foundation, Athens, GR-11635, Greece*

[a]Corresponding author: nbacalis@eie.gr
[b]zhuangx@seu.edu.cn

**Abstract.** We demonstrate that, if a truncated expansion of a wave function is large, then the standard excited states computational method, of optimizing one "root" of a secular equation, according to the theorem of Hylleraas, Undheim and McDonald (HUM), tends to the correct excited wave function, comparable to that obtained via our proposed minimization principle for excited states [J. Comput. Meth. Sci. Eng. **8**, 277 (2008)] (independent of orthogonality to lower lying approximants). However, if a truncated expansion of a wave function is small - that would be desirable for large systems - then the HUM-based methods may lead to an incorrect wave function - despite the correct energy (: according to the HUM theorem) whereas our method leads to correct, reliable, albeit small truncated wave functions. The demonstration is done in He excited states, using truncated series "small" expansions both in Hylleraas coordinates, and via standard configuration-interaction truncated "small" expansions, in comparison with corresponding "large" expansions. Beyond that, we give some examples of linear combinations of Hamiltonian eigenfunctions that have the energy of the 1st excited state, albeit they are orthogonal to it, demonstrating that the correct energy is not a criterion of correctness of the wave function.

## Introduction

In *ab-initio* computations of the properties of any system the wave functions are necessarily truncated. Large wave function expansions in truncated - but as complete as possible - Hilbert spaces, are generally safe, but if the system is large, the large wave functions are rather impracticable. Thus, small and handy, but reliable, expansions are much preferable, provided that they curry the main properties of the system, leaving the improvement of the energy to any correlation-corrections (by describing the "splitting" of the wave function in areas of accumulated electrons). For the ground state, such a "useful" wave function is relatively easily obtained by minimizing the energy, but for excited states, minimization of the energy can only be achieved if the wave function is orthogonal to all lower states. However, this requires accurate large expansions; otherwise, orthogonality to inaccurate approximations of lower states *must* lead to an energy lying below the exact (without collapsing to lower states) [1]. On the other hand, if the calculation of the ground state and excited states are optimized by variation respectively, such as the State-Specific Theory [2], based on approximated orthogonality, the energy of the ground state and the excited states can be obtained with good accuracy, but the orthogonality between the whole ground and excited state wave functions will be destroyed, and thus can not guarantee the accuracy of the approximate *wave function*. Particularly, this would render the calculation of properties between different states (e.g., optical transitions, the oscillator strength, etc.) unable to guarantee its reliability. This is why the spectral line *positions* of atoms and ions, obtained by many experiments and astronomical observations can be accurately theoretically explained, but their intensity distribution is difficult to be obtained with satisfactory theoretical description.

Here we compute both "large" and "small" wave functions of ground and excited states of He (the smallest non-trivial system) using two methods, the standard method based on the theorem of Hylleraas, Undheim and McDonald (HUM) [3] of optimizing desired "roots" of a secular equation, and our method, based on our proposed minimization principle for excited states (VPES, "*F*") [1], which - it is important to mention - *does not use any orthogonality to lower-lying approximate* wave functions. This allows approaching the exact Hamiltonian eigenfunction, even in small truncated spaces. For demonstration purposes we use two kinds of wave function expansions, i.e. (i) series expansions in Hylleraas coordinates [3] (the most accurate for He atom) and (ii) configuration interaction (CI) expansions in standard spatial coordinates [1].

First we demonstrate that if the expansion is large enough, then - as expected for large expansions - the standard methods based on HUM, and our method, "*F*", give practically identical wave functions, confirming that our method is valid and reliable.

Further, considering the traditional standard method of orthogonalizing to all *truncated* wave functions lying lower than the desired excited one, we demonstrate that if the functions are "small" - i.e., if, in optimizing the desired root of the (small) secular equation, the optimized small excited function is orthogonal to the lower roots that are *small* functions - then, in spite of the fact that the energy may converge from above to the correct value according to the theorem of Hylleraas, Undheim and McDonald (HUM) [3], the imposed orthogonality (i.e. to *truncated approximants*) may lead to disastrously incorrect *main* orbitals which are unable to describe the main properties, and indispensably need correlation orbitals that try to "improve" the *total* wave function (not the *main* orbitals), thus prohibiting a correct identification of the *main* orbitals, e.g. as to a correct "HOMO" or "LUMO". On the contrary, minimization of our proposed functional *F* [1], leads to a correct (small) wave function, which curries the same *main* properties as the "large" function (obtained comparable, and safely, by either HUM or *F*). As an example, in He $^3S$ 1$s$3$s$, HUM "small" leads to 1$s$**2**$s$+*correlation corrections* (instead of 1$s$3$s$), whereas our "*F*" "small" leads to the correct 1$s$3$s$. It is important to note that *F* has a local minimum *at* the excited state, therefore, it does **not** need any orthogonalization to lower lying wave functions, it just insures solution of the Schrödinger equation; orthogonality to exact lower states should be an outcome.

The demonstration is done in He excited states, using truncated series expansions in Hylleraas coordinates, as well as standard configuration-interaction (CI) truncated expansions in comparison with corresponding "large" expansions.

In the following, after a brief discussion about the saddleness of the excited state energy, we present:

(1) Our tools and approximations;
(2) Our methods;
(3) We demonstrate:

(i) The validity and reliability of our method by showing the equivalence of "large" expansions obtained either via HUM or via *F*, i.e. by displaying the *main* orbitals, as well as average values of $\langle r^n \rangle$, $n$=-1,0,1,2;

And:

(ii) That "small" expansions obtained via *F* are correct (with *main* orbitals comparable to those of "large" functions) whereas those obtained via HUM may give misleadingly incorrect

*main* orbitals, with concomitant incorrect practical conclusions as to which HUM orbitals are HOMO/LUMO, or as to which HUM orbitals contribute to charge transfer.

(4) Then, since the excited states are saddle points [cf. below], and around a saddle point some actual computation, like minimizing "$F$", may stop, to within a convergence criterion, at the side with energy of either $E+\Delta E$ or $E-\Delta E$, we state three reliability criteria in case that the minimization converges at the side of energy just *below* the saddle point energy of the correct excited state, i.e. at $E-\Delta E$ - a subject which is never encountered by any standard method based on HUM because HUM always converges from above (to a function [cf. function "$\Phi_{HUM}$" in Fig. 1] which is *necessarily veered away* from the *exact eigenfunction* [cf. function "$\psi_1$" in Fig. 1]).

(5) Finally we give some examples of truncated wave functions that have exactly the energy of an excited state, but are *orthogonal* to it(!), demonstrating that the correct energy is not a safe criterion of the correctness of the wave function.

## The saddleness of the excited state energy $E_n$

Expand the approximant $|\phi_n\rangle$ around the exact state $|n\rangle$ (assumed real, non-degenerate, and normalized) in terms of the exact Hamiltonian eigenfunctions $|i\rangle$, $i=0,1,...n,...$, with energies $E_0 < E_1 < E_2 \cdots$

$$|\phi_n\rangle = \sum_{i<n}|i\rangle\langle i|\phi_n\rangle + |n\rangle\sqrt{1-\sum_{i\neq n}\langle i|\phi_n\rangle^2} + \sum_{i>n}|i\rangle\langle i|\phi_n\rangle \tag{1}$$

The energy is written as $\langle \phi_n|H|\phi_n\rangle = E = -L + E_n + U$ where, in terms of the coefficients, the lower terms, $L$, and the higher terms, $U$, (saddle) are:

$$\begin{array}{ccc} \textbf{\textit{down - paraboloids}} & & \textbf{\textit{up - paraboloids}}: \\ E = \boxed{-L = -\sum_{i<n}(E_n-E_i)\langle i|\phi_n\rangle^2} + E_n + \boxed{U = \sum_{i>n}(E_i-E_n)\langle i|\phi_n\rangle^2} \end{array} \tag{2}$$

If $L$ is absent, i.e. if $n=0$ (Eckart theorem [4]), or if $\phi_n$ is orthogonal to all lower $|\psi_i\rangle = |i\rangle$ (which can be approximated by $\phi_i$ satisfactorily only if $\phi_i$ are "large" expansions), then minimizing $E=E_n+U$ is sufficient. But if $\phi_i$ are "small" expansions (not accidentally orthogonal to $\phi_n$), then $L$ is present, and $E$ has a saddle point at $|n\rangle$. Then, minimizing $E=-L+E_n+U$ "orthogonally to all lower approximants $\phi_i$", must lead to $E$ below $E_n$; because (consider e.g. the 1$^{st}$ excited state): If $\phi_0$ is such an (inaccurate) approximant of $\psi_0$, i.e. if $\phi_0$ is not orthogonal to $\psi_1$, then in the subspace orthogonal to $\phi_0$, there is a function $\phi_1^+$, which is: "**closest** to $\psi_1$ (**while** orthogonal to $\phi_0$)". Closest to $\psi_1$ means it has no other components out of the plane of $\{\phi_0, \psi_1\}$. It is easily seen that this function, $\phi_1^+$, lies **below** $E_1$ (therefore, the minimum lies even lower):

$$\phi_1^+ \equiv (\psi_1 - \phi_0\langle\psi_1|\phi_0\rangle)/\sqrt{1-\langle\psi_1|\phi_0\rangle^2} \Rightarrow$$
$$\langle\phi_1^+|H|\phi_1^+\rangle = E[\phi_1^+] = E[\psi_1] - (E[\psi_1]-E[\phi_0])\langle\psi_1|\phi_0\rangle^2/(1-\langle\psi_1|\phi_0\rangle^2) \Rightarrow \tag{3}$$
$$E[\phi_1^+] < E[\psi_1] = E_1$$

But the HUM theorem demands $E[\phi_n] > E_n$ while $\phi_n$ ***is*** orthogonal to all lower "roots" $\phi_i$ of the secular equation. Therefore, on the subspace orthogonal to the *lowest root*, the function which is **closest** to $\psi_1$, **while** orthogonal to the lowest root, i.e. $\phi_1^+$, is not accessible by HUM (and even more inaccessible is $\psi_1$ itself). Even worse, note that in optimizing any HUM root (say $\phi_1$), all other roots ($\phi_0, \phi_2, ...$) get deteriorated since we may have $\langle 1|\phi_1\rangle^2 \to 1$, *at will*, *but*:

$$\langle 0|\phi_0\rangle^2 + \langle 0|\phi_1\rangle^2 + \langle 0|\phi_2\rangle^2 + \cdots + \langle 0|\phi_N\rangle^2 \leq 1 \Rightarrow \langle 0|\phi_0\rangle^2 < 1 - \langle 0|\phi_1\rangle^2 \text{ (: accuracy upper bound)} \quad (4)$$

This means that, if $\langle 1|\phi_1\rangle^2$ approaches $\to 1$, *at will*, then $\langle 0|\phi_0\rangle^2$ *cannot approach* $\to 1$, *at will*, $\phi_0$ having to be deteriorated (of worse quality than $\phi_1$), especially if it is "small" because then, $\langle 1|\phi_1\rangle^2 < 1$, $\langle 0|\phi_1\rangle^2 < 1$.

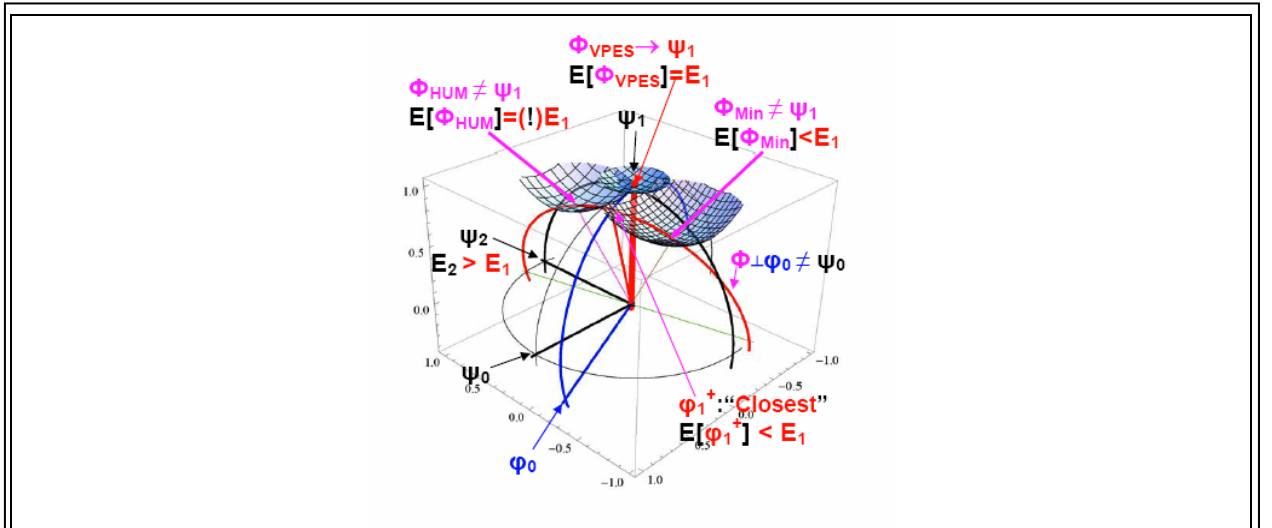

Fig. 1. Schematic representation of states. All states are assumed normalized: $\psi_0$, $\psi_1$, $\psi_2$, ... with energies $E_0 < E_1 < E_2 < ...$ are the unknown exact eigenstates; $\varphi_0$ is a known approximant of $\psi_0$ (not orthogonal to $\psi_1$). The subspace orthogonal to $\varphi_0$, is S = $\{\Phi, \varphi_1^+\}$ (oblique circle orthogonal to the vertical circle of $\{\varphi_0, \psi_1\}$ and orthogonal to $\varphi_0$), and if $\varphi_0$ is not (accidentally) orthogonal to the unknown $\psi_1$, then S = $\{\Phi, \varphi_1^+\}$ does not contain $\psi_1$. In the subspace S = $\{\Phi, \varphi_1^+\}$, the closest approximant to $\psi_1$ is $\varphi_1^+$ (the intersection of the oblique circle and the vertical circle of $\{\varphi_0, \psi_1\}$), and, as explained in the text, $\varphi_1^+$ lies below $\psi_1$: $E[\varphi_1^+] < E_1$. In going, in S, orthogonally to $\varphi_0$, from $\varphi_1^+$ toward a state near $\psi_2$ (the diameter of the oblique circle near $\psi_2$), i.e. in going, in S, from $E[\varphi_1^+] < E_1$ toward $E \approx E_2 > E_1$, one passes from $E_1$, i.e. from states, "$\varphi_1$", of S, orthogonal to $\varphi_0$, but having energy $E[\varphi_1] = E_1$. If, in optimizing $\varphi_1$ by HUM theorem, $\varphi_0$ is the lowest (deteriorated, as explained in the text) root of the secular equation, then the 2nd "root", $\varphi_1 = \Phi_{HUM}$, is one of these states, "$\varphi_1$", with lowest possible (allowed by HUM theorem) energy $E[\varphi_1] = E[\Phi_{HUM}] = E_1$ (cf. left "minimum" in the figure on the oblique circle S). But, evidently, $\varphi_1 = \Phi_{HUM}$ is not $\psi_1$. It might be desirable to continue optimization in S toward, at least, $\varphi_1^+$, the closest, in S, to $\psi_1$. But HUM theorem prohibits such a continuation, since the 2nd root must always be *higher* than $E_1$. In an attempt to approach, as much as possible, $\psi_1$, one might try, by other means, i.e. by direct minimization, to minimize the energy, in S, orthogonally to $\varphi_0$, toward $\varphi_1^+$. But $\varphi_1^+$ is not a critical point, and the minimum in S, orthogonal to $\varphi_0$, lies even lower: $E[\Phi_{Min}] < E_1$ (cf. right "minimum" in the figure, also on the oblique circle S). $\Phi_{Min}$ does not suffer from variational collapse, since it is orthogonal to $\varphi_0$. $\Phi_{Min}$ is not a "bad" approximant of $\psi_0$, it is rather an approximant of $\psi_1$, probably as good (or as bad) as $\Phi_{HUM}$. Both $\Phi_{HUM}$ and $\Phi_{Min}$ are veered away from $\varphi_1^+$, in S, and, therefore, from $\psi_1$. On the other hand, "F", the reported "Variational Principle for Excited States" (VPES) approaches $\psi_1$, (in general the exact excited states $\psi_n$), $\Phi_{VPES} \to \psi_1$, *independently* of orthogonality to $\varphi_0$ (to lower lying approximants), and regardless of the accuracy of the latter, i.e. of their closeness to the exact, provided that the lower approximants, used in VPES, are reasonable approximants, as explained in the text (cf. the upper "minimum" in the figure at $\psi_1$). If the parameter space is "large", then $\varphi_0$ tends to $\psi_0$, the oblique circle (orthogonal to $\varphi_0$) tends to rise, $\varphi_1^+$ tends to $\psi_1$, and the three "minima" tend to coincide at $\psi_1$.

Therefore, the *optimized* HUM 2nd root $\phi_1$ (although $E[\phi_1] > E_1 > E[\phi_1^+]$) is orthogonal to a **deteriorated** 1st root $\phi_0$, which, consequently, has a **deteriorated** $\phi_1^+$ (the **"closest to $\psi_1$ while** orthogonal to the **deteriorated $\phi_0$"**) (cf. Fig. 1); i.e. $\phi_1$, moving in space orthogonal to **deteriorated $\phi_0$s**, just stops at $E_1$ and cannot approach a **deteriorated $\phi_1^+$**, thus, the *optimized* HUM 2nd root **$\phi_1$ is much more veered away from the exact $\psi_1$**. This is clearly demonstrated below for He.

(If the optimized wave functions, as HUM roots, are misleading even for the smallest atom, He, then **there is no guarantee for the correctness of HUM roots in larger systems!**)

## Tools and approximations

First we need very accurate (truncated) functions $\Psi_n$ *to resemble* $\simeq$ eigenfunctions $\psi_n$ as well as truncated approximants $\Phi_n$ to check the closeness to $\psi_n$ i.e. to $\Psi_n$.

As truncated functions we use:
1. For He $^1S$ ($1s^2$ and $1s2s$): **Series expansion in Hylleraas variables** $s = r_1 + r_2$, $t = r_1 - r_2$, $u = |\vec{r}_1 - \vec{r}_2|$.

The two-electron wave function $\Phi(r_1, r_2)$ consist of one **Slater determinant** of variational Laguerre–type orbitals (VLTOs), $1s$, $2s$

$$\chi(n, r; z_n, \{a_{n,k}\}) = \frac{4\sqrt{\pi}\sqrt{(n-1)!n!}}{n^2} z_n^{3/2} \sum_{k=0}^{n-1} \frac{a_{n,k}(-2rz_n/n)^k e^{-rz_n/n}}{k!(k+1)!(n-k-1)!} \quad (5)$$

(where $r_1 = (s+t)/2$, $r_2 = (s-t)/2$ and $a_{n,k}$, $z_n$ are free variational parameters), and where the determinant is **multiplied** by a truncated power series of $s, t, u$:

$$\Phi(r_1, r_2) = Det|\chi_1, \chi_2| \times \sum_{i_s, i_t, i_u = 0}^{n_s, n_t, n_u} c_{i_s, i_t, i_u} s^{i_s} t^{2i_t} u^{i_u}, \quad (6)$$

where $c$'s are linear variational parameters, comprising the eigenvectors of the 1$^{st}$ or 2$^{nd}$ root of a secular equation.

For the "exact" $\Psi_n$ we take terms up to $(n_s, n_t, n_u) =$ **(2,2,2): 27 terms**, $E_0 \approx$ -2.90371 a.u., $E_1 \approx$ -2.14584 a.u., compared to Pekeris' 95 terms: $E_0 =$ -2.90372, $E_1 =$ -2.14597 a.u. [5]. For the "truncated" trial "small" functions $\Phi_n$ we take terms up to $(n_s, n_t, n_u) =$ **(1,1,1): 8 terms.**

2. For He $^1S$ ($1s^2$, $1s2s$ and $1s3s$) and for He $^3S$ ($1s2s$ and $1s3s$) we use **Configuration Interaction (CI) in standard spherical coordinates** $(r, \theta, \varphi)$.

In this approximation the two-electron wave function $\Phi(r_1, r_2)$ is a linear combination of configurations composed of Slater determinants (SD) of atomic variationally optimized Laguerre-type (VLTO) spin-orbitals, which have been proven [6] to provide conciseness, clear physical interpretation and near equivalent accuracy with *numerical* multi-configuration self-consistent field (MCHF) - which is one of the most accurate numerical methods for atomic CI calculations. The VLTO atomic orbitals are orthogonalized by appropriate $g_k^{n,\ell}$ factors:

$$A^{n,\ell,m}\left(\sum_{k=0}^{n-\ell-1} c_k^{n,\ell} g_k^{n,\ell} r^{(\ell+k)} e^{-z_{n,\ell}\frac{r}{n}} + b^{n,\ell} e^{-q^{n,\ell} z_{n,\ell}\frac{r}{n}} \delta_{\ell,0}\right) Y^{\ell,m}(\theta, \phi) \quad (7)$$

where, variational parameters are: the quantities $z_{n,\ell}$, $b^{n,\ell}$, $q^{n,\ell}$, along with the linear CI coefficients as eigenvectors of the roots of the secular equation.

As "exact" $\Psi_n$ we use a "large" expansion in **1s, 2s, 3s, 4s, 5s, 2p, 3p, 4p, 5p, 3d, 4d, 5d, 4f, 5f**. For $^1S$: $E_0 \approx$ -2.90324 a.u., $E_1 \approx$ -2.14594 a.u., $E_2 \approx$ -2.06125 a.u. (exact: -2.06127 a.u. [5]), for $^3S$: $E_0 \approx$ -2.17521 a.u., $E_1 \approx$ -2.06869 a.u. (exact: -2.17536, -2.06881 a.u. [5]). (The ground state is slightly harder to converge: it needs more configurations near the nucleus.) As "truncated" trial functions $\Phi_n$ we use a "small" expansion in **1s, 2s, 3s**.

## The two methods used

In both of the above approximations we shall use **two methods**:
1. Minimimizing (optimizing) directly the $n^{th}$ **HUM root**, which, as aforementioned, **must** be veered away from the exact eigenfunction $\psi_n$ (because, according to the HUM theorem it tends to the exact energy **from above**, i.e. it cannot go to **lower energies**, required in order to approach closer the exact $\psi_n$, due to orthogonality to lower roots **which are deteriorated**).

2. Minimimizing the functional $F_n$ that has **minimum <u>at</u>** the **exact $\psi_n$** [1]

$$F_n[\phi_0,\phi_1,...;\phi_n] \equiv E[\phi_n] + 2\sum_{i<n}\frac{\langle\phi_i|H-E[\phi_n]|\phi_n\rangle^2}{E[\phi_n]-E[\phi_i]}\left[1-\sum_{i<n}\langle\phi_i|\phi_n\rangle^2\right]^{-1}. \tag{8}$$

$F_n$ is obtained by inverting the sign of $L$ (the down parabolas): $E = +L + E_n + U$ in Eq.2. The lower $\phi_i$ are allowed to be rather **inaccurate and very "small"**, provided that the Hessian and all its principal minors along the main diagonal be positive (cf. Sylvester theorem), which is easily fulfilled because their main term in equations 9, 10 below (cf. ref. [1]) is large and the overlaps in $1+2[O(\text{coefficients}^2)]$ in equations 9, 10 are small: Indeed, the Hessian, and all its principal minors along the main diagonal, are, respectively, of the form:

Hessian:

$$A_n^n = 2^{n+1}\prod_{i=0}^{n-1}(E_n-E_i)\left(E[\phi_n^{\perp\{n\}}]-E_n\right)\times\left(1+2\left[O\left(\langle n|\phi_i\rangle^2\right)\right]\right) \tag{9}$$

and its principal minors along the main diagonal:

$$A_n^{k<n} = 2^{k+1}\prod_{i=0}^{k}(E_n-E_i)\times\left(1+2\left[O\left(\langle n|\phi_i\rangle^2\right)\right]\right) \tag{10}$$

We emphasize once again that the functional $F_n$ *does not use any orthogonality to lower-lying approximate* wave functions. This allows approaching the exact Hamiltonian eigenfunction, even in small truncated spaces.

The procedure of computing and minimizing $F_n$ is shown in the logical diagram of Fig. 2.

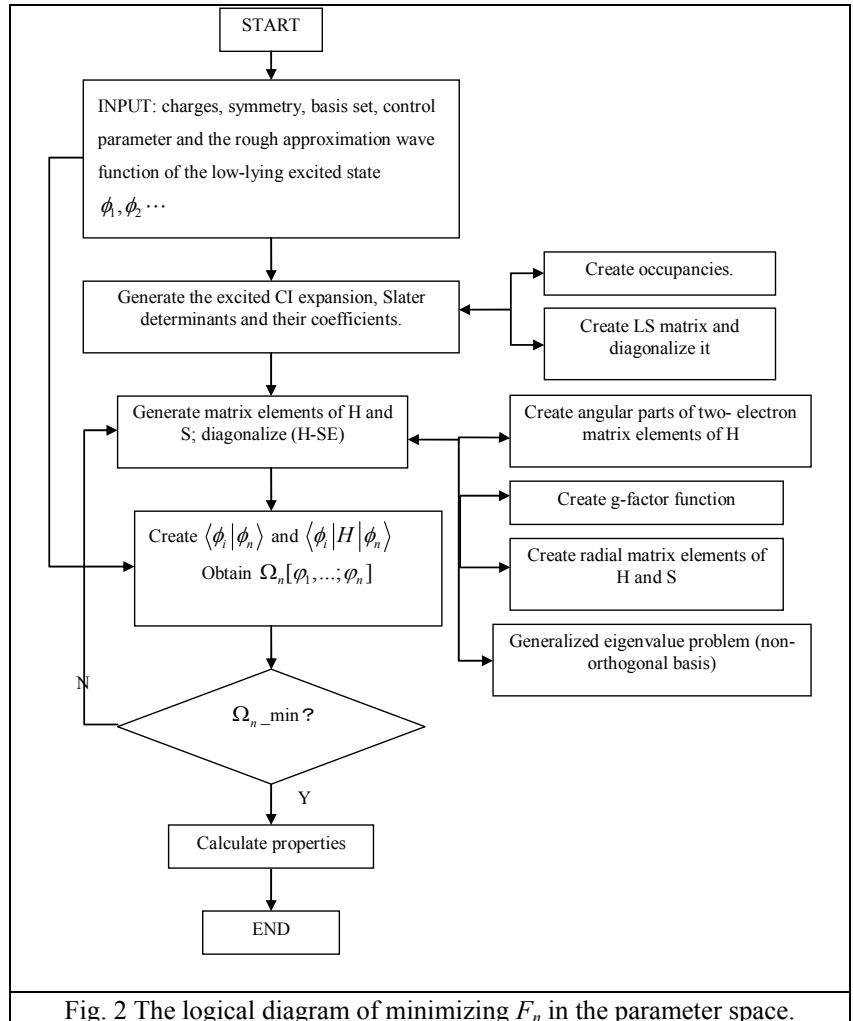

Fig. 2 The logical diagram of minimizing $F_n$ in the parameter space.

## Results

1. In Hylleraas coordinates the discrepancy between "*F*"- and HUM is clear: The **main** orbitals of the Hylleraas wave functions for **He $^1S$ 1$s$2$s$** are shown in Fig. 3. Clearly, the HUM-wave function is **not 1$s$2$s$**! (**"2$s$"** has a "node" at 10 a.u.! Therefore, **"2$s$"** is essentially 1$s'$ -which makes the state 1$s$1$s'$). However, the **total** wave function, to be improved, needs **8** Hylleraas series **terms** (or even **27** terms!). On the contrary, the "*F*"- wave function (both the "large", of 27 terms, and the "small", of 8 terms, are correct and practically identical. This was expected because $F_1$ has minimum **at** the exact (saddle point) 1$s$2$s$, whereas the HUM 2$^{nd}$ root is orthogonal to a necessarily deteriorated 1$^{st}$ root [cf. Fig.1], therefore it is veered away from the exact 1$s$2$s$.

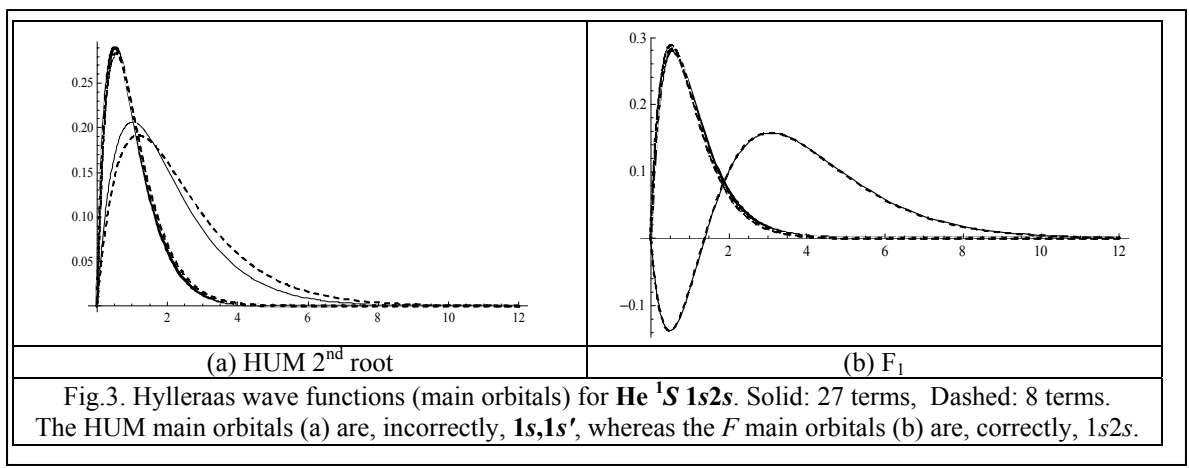

| (a) HUM 2$^{nd}$ root | (b) $F_1$ |

Fig.3. Hylleraas wave functions (main orbitals) for **He $^1S$ 1$s$2$s$**. Solid: 27 terms, Dashed: 8 terms. The HUM main orbitals (a) are, incorrectly, **1$s$,1$s'$**, whereas the *F* main orbitals (b) are, correctly, 1$s$2$s$.

2. Using the CI expansion in standard spherical coordinates first we establish the validity of our variation principle for excited states VPES, "*F*", by comparing our "*F*-large" wave functions with standard "HUM-large" functions.

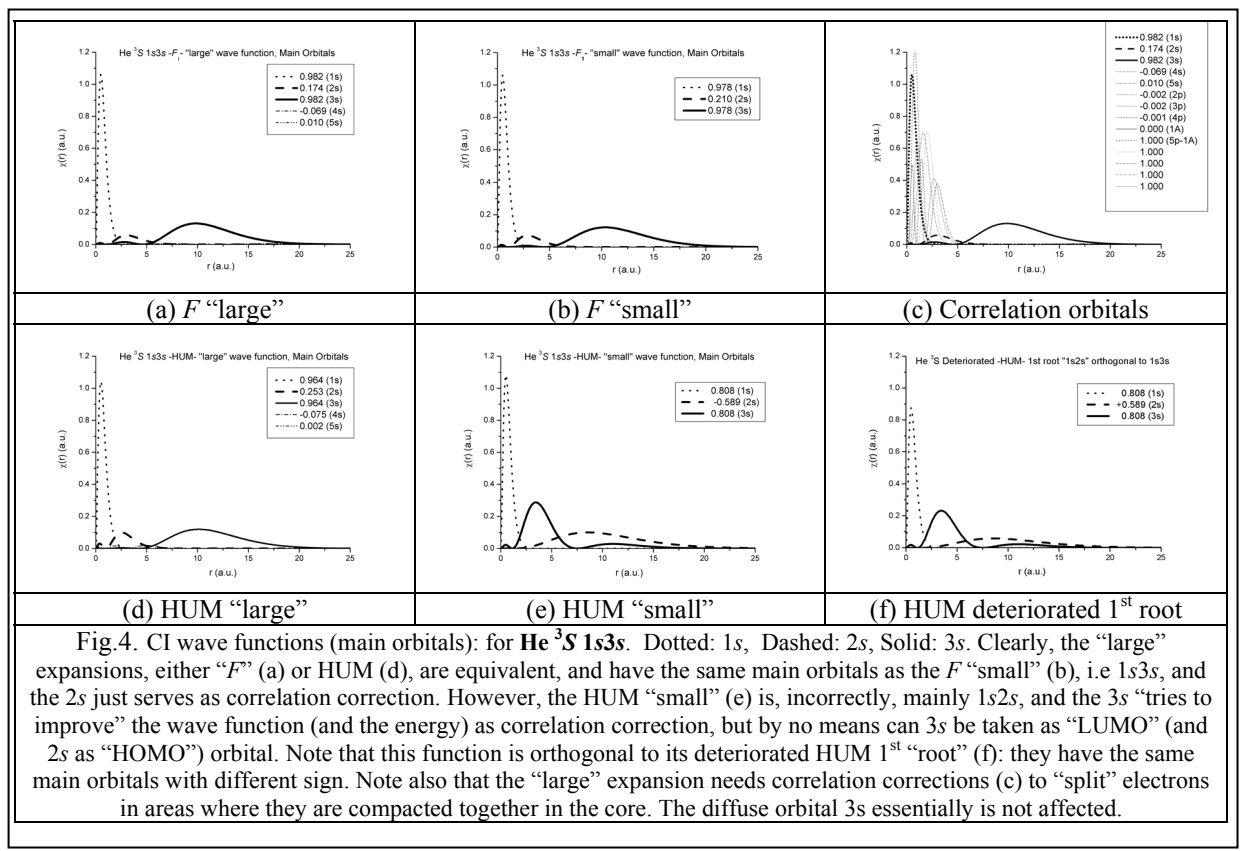

| (a) *F* "large" | (b) *F* "small" | (c) Correlation orbitals |
| (d) HUM "large" | (e) HUM "small" | (f) HUM deteriorated 1$^{st}$ root |

Fig.4. CI wave functions (main orbitals): for **He $^3S$ 1$s$3$s$**. Dotted: 1$s$, Dashed: 2$s$, Solid: 3$s$. Clearly, the "large" expansions, either "*F*" (a) or HUM (d), are equivalent, and have the same main orbitals as the *F* "small" (b), i.e 1$s$3$s$, and the 2$s$ just serves as correlation correction. However, the HUM "small" (e) is, incorrectly, mainly 1$s$2$s$, and the 3$s$ "tries to improve" the wave function (and the energy) as correlation correction, but by no means can 3$s$ be taken as "LUMO" (and 2$s$ as "HOMO") orbital. Note that this function is orthogonal to its deteriorated HUM 1$^{st}$ "root" (f): they have the same main orbitals with different sign. Note also that the "large" expansion needs correlation corrections (c) to "split" electrons in areas where they are compacted together in the core. The diffuse orbital 3$s$ essentially is not affected.

In Fig. 4 the CI wave functions (main orbitals), HUM and $F$, are compared for the triplet **He $^3S$ 1s3s**. The "large" functions $F$ (a) and HUM (d) are almost identical. Notice that the $F$ "small" function (b) has the same main orbitals as the "large" functions $F$ (a) and HUM (d), namely 1s and 3s, where the 2s just adds some correlation correction near the nucleus (as well as all higher orbitals of the "large" expansion (c)). On the contrary, the HUM "small" expansion (e), which is orthogonal to a deteriorated 1$^{st}$ root $^3S$ "1s2s" (f), **has main orbitals 1s2s** (with opposite sign), while the 3s orbital, as correlation correction, tries to correct the total wave function and approach the correct energy. Thus, the HUM solution proposes to the audience, as "HOMO" orbital, the 2s instead of the 3s, therefore, it **is misleading**. Of course, blindly considering as "LUMO" the 1$^{st}$ higher "unoccupied" orbital, i.e. the HUM 3s, is completely out of question.

Similar results are obtained also for the singlet **He $^1S$ 1s3s**: $F$ "small" is, correctly, mainly 1s3s, whereas HUM "small" is, **misleadingly**, mainly 1s2s, instead of 1s3s [cf. Fig. 5]. Notice that the lowest unoccupied orbital, "LUMO", is just a correlation orbital, trying to improve the *total* wave function *near the nucleus*, and by no means should it be considered as the first candidate orbital to be occupied by an excited electron.

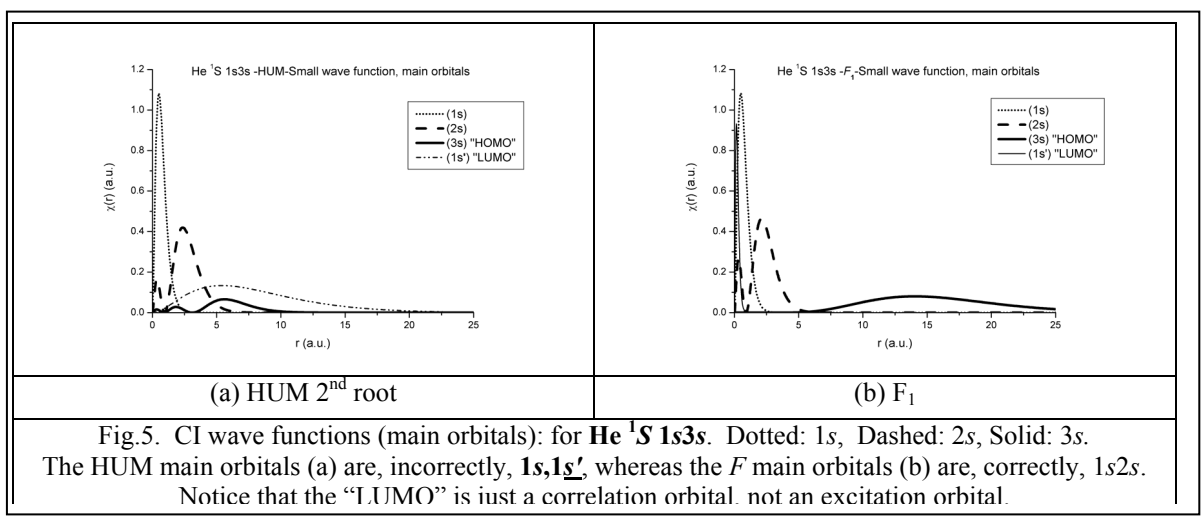

Fig.5. CI wave functions (main orbitals): for **He $^1S$ 1s3s**. Dotted: 1s, Dashed: 2s, Solid: 3s. The HUM main orbitals (a) are, incorrectly, **1s,1s′**, whereas the $F$ main orbitals (b) are, correctly, 1s2s. Notice that the "LUMO" is just a correlation orbital, not an excitation orbital.

Nevertheless, for He $^1S$ 1s2s the "small" HUM and $F$ functions are the essentially the same.

3. The reliability of our method is also seen by the expectation values of $\langle r^n \rangle$, $n$=-1,1,2 in comparison of triplet He $^3S$ states with the literature. In order to demonstrate the quality of our $F$-"large" atomic wave functions, we use the same VLTO orbitals up to 5f, configurations (Cs) and Slater determinants (SDs) in our calculations by minimizing the $F_n$ functional. We use, as standard reference, the most reliable calculated atomic data obtained by Chen M K (1994) [7] by B-splines method using more than 100 B-spline functions. We demonstrate the quality of our atomic wave functions (beyond Fig.4) by calculating their relative errors respectively.

In Table 1, the first column shows the variation results obtained by optimizing the lowest root of the secular equation via $F_n$ minimization. The $2^3S$ wave function was obtained at its energy minimum, following the Eckart Theorem [4], by using 14 VLTOs (1s, 2s, 3s, 4s, 5s, 3p, 4p, 5p, 3d, 4d, 5d, 4f, 5f), full CI with 20 Cs, 50 SDs. We calculated the wave function $3^3S$ at the minimum of $F_1$, by using 13 VLTOs (1s, 3s, 4s, 5s, 2p, 3p, 4p, 5p, 3d, 4d, 5d, 4f, 5f), full CI with 16 (Cs), 46 Slater determinants (SDs) where the first order approximate wave function $\phi_0$ was calculated with 2 VLTOs (1s,2s), 1 Cs, 1 SD, $E[\phi_0]$ = -2.1718648 a.u.. The wave function $4^3S$ was calculated at the minimum of $F_2$, by using 12 VLTOs (1s, 4s, 5s, 2p, 3p, 4p, 5p, 3d, 4d, 5d, 4f, 5f), full CI with 13 (Cs), 43 Slater determinants (SDs), using the above $\phi_0$ (1s,2s), while the first order approximate wave function $\phi_1$ was calculated with 2 VLTOs (1s,3s), 1 Cs, 1 SD, $E[\phi_1]=-2.0685536$. The third column shows reference values. In the fourth column the relative errors compared with reference C are given. In the last column the latest MCHF standard values are given (cf. http://nlte.nist.gov/MCHF/index.html, National Institute of Standards and Technology (NIST) MCHF Database).

Table 1. The energy values and radial expectation values of the three lowest lying states of He $^3S^{(e)}$ in atomic units.

|       |         | F-"Large" | Chen M K   | Relative error     | Exact      |
|-------|---------|-----------|------------|--------------------|------------|
|       |         | F         | C          | \|1-F/C\|          |            |
| $2^3S$ | $E_1$   | -2.1752135 | -2.1752288 | 7.033 $10^{-06}$  | -2.1753598 |
|       | $<1/r>$ | 1.15465   | 1.154664   | 3.464 $10^{-06}$  |            |
|       | $<r>$   | 2.55057   | 2.550468   | 3.999 $10^{-05}$  |            |
|       | $<r^2>$ | 11.46337  | 11.46438   | 8.810 $10^{-05}$  |            |
| $3^3S$ | $E_2$   | -2.0686826 | -2.0686888 | 2.997 $10^{-06}$ | -2.0688059 |
|       | $<1/r>$ | 1.06361   | 1.063674   | 6.017 $10^{-05}$  |            |
|       | $<r>$   | 5.85684   | 5.855982   | 1.465 $10^{-04}$  |            |
|       | $<r^2>$ | 68.74368  | 68.70871   | 5.090 $10^{-04}$  |            |
| $4^3S$ | $E_3$   | -2.0364988 | -2.0365120 | 6.482 $10^{-06}$ | -2.0366259 |
|       | $<1/r>$ | 1.03462   | 1.034570   | 4.833 $10^{-05}$  |            |
|       | $<r>$   | 10.67651  | 10.66123   | 1.433 $10^{-03}$  |            |
|       | $<r^2>$ | 239.60138 | 238.580    | 4.042 $10^{-03}$  |            |

As seen from the comparison with Chen [7], our "Large" functions are indeed reliable (see also Fig. 4). However, for the first excited state $3^3S$, the F "Small" energy, $E_1$=-2.0693402 a.u. is lower than the MCHF standard value (-2.0688059 a.u.), while, of course, the $F_1$ value itself is a little higher (-2.0679501 a.u.). In other words, in stopping the $F_1$ minimization, it happened that the convergence criterion was satisfied *near* the exact saddle point, in the side of $E-\Delta E$. Below we establish a criterion of correctness of the wave function, independent of the literature, based on the augmentation $\delta^{(e)}$ of the energy toward the unknown exact eigenvalue. Similarly for the second excited state $4^3S$ using the same CI "small" expansions, the wave function obtained by the F functional has energy value $E_2$=-2.0366305 a.u. i.e. lower than the MCHF standard value (-2.0366259 a.u.), while the $F_2$ value itself is, of course, a little higher (-2.0363067 a.u.). Therefore, it is demonstrated that for the excited state, the lower lying wave function would be trustable provided that the augmentation $\delta^{(e)}$ toward the exact value (cf. below) is small. Using our wave functions, our $\delta^{(e)}$ augmentation values are $\delta^{(2)}$ = 6.819 x $10^{-4}$ a.u. for the $3^3S$ state, and $\delta^{(3)}$= 2.053 x $10^{-4}$ a.u. for the $4^3S$ state.

### Reliability Criteria

The **exact** $E_n$ is a **saddle point** [cf. Fig. 6] so, around it there are points with either **slightly higher** or **slightly lower** energy. Then the question arises: If we stop the F minimization, by fulfilling some convergence criterion, and it happens that the energy be either **slightly lower** (or slightly higher), then is the "small" wave function reliable?

There are two criteria: (i) In the parameter space around the F minimum we must check that the final point (minimum of F) is indeed saddle in energy E [cf. Fig. 6]. (ii) Since this will never happen *exactly*, the difference of both known values (F-E) must be check to be small. If a "large" function is available (to serve as "exact"), before we release the "small" function to the audience, there is a third criterion: In $E = -L + E_n + U$ [cf. Eq.2], the unknown U is $U \geq 0$.

Then, $E \geq -L + E_n$. In fact, we should notice that **this** is **the correct lower bound** of $E$, and **not just** $E_n$, because $E_n$ would be a lower bound of $E[\phi_n] > E_n$ if either (a) $\phi_n$ were exactly orthogonal to all lower eigenfunctions (which **never** happens and is approximately fulfilled if the functions are "large" expansions) or (b) if $\phi_n$ were the $(n+1)^{th}$ HUM root (which is **always veered away** from the exact eigenfunction (cf. Fig.1) and approaches the exact only if it also is a "large" expansion). But Shull and Löwdin [8] have shown that any excited state can be computed independently of the lower lying approximants, and this, exactly, is done by $F$. Therefore, the **correct** lower bound is

$$E = E[\phi_n] \geq -L + E_n = -\sum_{i<n}(E_n - E_i)\langle i|\phi_n\rangle^2 + E_n \quad (9)$$

so that the 3$^{rd}$ reliability criterion is that (iii) $0 < E_n - E \leq L = \delta^{(n)}$, where $L$ should be small.

Using as $\psi_0$ our "large" wave functions $\phi_0$, we obtained Table 2. We have two cases **below** the exact. But $F - E$ is small, (YES) $L > E_{exact} - E$, and $L$ is small, so, they are reliable.

Table 2. Estimating the 3$^{rd}$ reliability criterion.

| Wave function | $F$ | $E$ | $F - E$ | $E_n$ | $E - E_n$ | $L$ | $L > E_n - E$ ? |
|---|---|---|---|---|---|---|---|
| $^1S$ $1s2s$ Large | -2.145934 | -2.14594 | 2 10$^{-9}$ | -2.14597 | 3 10$^{-5}$ | 2 10$^{-9}$ | |
| $^1S$ $1s2s$ Small | --- | --- | --- | --- | --- | --- | --- |
| $^1S$ $1s3s$ Large | -2.061252 | -2.061252 | 3 10$^{-13}$ | -2.06127 | 2 10$^{-5}$ | 3 10$^{-8}$ | |
| $^1S$ $1s3s$ Small | -2.049335 | -2.06278 | 1.3 10$^{-3}$ | -2.06127 | -0.002 | 0.006 | YES |
| $^3S$ $1s3s$ Large | -2.068272 | -2.068683 | 4.1 10$^{-4}$ | -2.06881 | 1.2 10$^{-5}$ | 5 10$^{-6}$ | |
| $^3S$ $1s3s$ Small | -2.067950 | -2.069361 | 1.4 10$^{-3}$ | -2.06881 | -5.3 10$^{-4}$ | 6 10$^{-4}$ | YES |
| $^3S$ $1s4s$ Large | -2.036494 | -2.036499 | 4.6 10$^{-7}$ | -2.03663 | 1.3 10$^{-4}$ | 7 10$^{-4}$ | |
| $^3S$ $1s4s$ Small | -2.036307 | -2.036630 | 3.2 10$^{-4}$ | -2.036626 | -4 10$^{-6}$ | 2 10$^{-4}$ | YES |

**Incorrect functions with "correct" energy**

Below we demonstrate that the correct energy is not a safe criterion of correctness of the wave function, since infinitely many functions $\Phi$ **orthogonal to $\psi_1$** can have the exact $E_1$ energy $E[\Phi] = E[\psi_1]$: From any function $\Psi$ obtain a normalized function orthogonal to $\psi_0$, $\psi_1$:

$$\Psi_\perp = \frac{\Psi - \psi_0\langle\psi_0|\Psi\rangle - \psi_1\langle\psi_1|\Psi\rangle}{\sqrt{1 - \langle\psi_1|\Psi\rangle^2 - \langle\psi_1|\Psi\rangle^2}} \quad (10)$$

and construct the following linear combination:

$$\Phi = \sqrt{\frac{E[\Psi_\perp] - E_1}{E[\Psi_\perp] - E_0}}\psi_0 + 0\psi_1 - \sqrt{\frac{E_1 - E_0}{E[\Psi_\perp] - E_0}}\Psi_\perp \quad (11)$$

Then, by construction, $\Phi$ has energy $E[\Phi] = E[\psi_1] = E_1$ **while it is orthogonal to $\psi_1$**. Some examples are shown in Table 3. Starting from a state $\Psi$ of energy $E[\Psi]$ we found a linear combination $\Phi = A\,\psi_0 + B\,\psi_1 + C\,\Psi$, such that: $<\psi_1|\Phi> \sim 0$, and $<\Phi|H|\Phi> = E[\psi_1] \sim E_1$.

Of course, these "$\Phi$" functions contain remote electrons and cannot be used as approximants of $\psi_1$, since they are orthogonal to $\psi_1$, so that the correct energy is not a safe criterion of correctness.

However, if entanglement is experimentally achieved, it might be possible to accomplish a chemical reaction at $E_1$ via more remote electrons.

| $E[\Psi]$ | $A$ | $B$ | $C$ | $<\psi_1|\Phi>$ | $<\Phi|H|\Phi>$ | $\psi_1$ | $E_1$ |
|---|---|---|---|---|---|---|---|
| -2.01990 | 0.37769 | 0.00004 | -0.92591 | -0.00002 | -2.14594 | $^1S$ | -2.14594 |
| -2.02117 | 0.37602 | -0.00006 | -0.92658 | -0.00002 | -2.14594 | $^1S$ | -2.14594 |
| -2.04555 | 0.34348 | -0.00120 | -0.93959 | -0.00128 | -2.14568 | $^1S$ | -2.14568 |
| -2.02260 | 0.55003 | 0.00124 | -0.83549 | 0.00007 | -2.06868 | $^3S$ | -2.06868 |
| -2.03650 | 0.47559 | -0.00012 | -0.87635 | 0.00006 | -2.06852 | $^3S$ | -2.06868 |
| -2.01990 | 0.58239 | 0.00020 | -0.81298 | -0.00040 | -2.05512 | $^1P$ | -2.05512 |
| -2.03106 | 0.50942 | -0.00063 | -0.86052 | -0.00035 | -2.05512 | $^1P$ | -2.05512 |
| -2.03223 | 0.50818 | 0.00442 | -0.86254 | 0.00104 | -2.05805 | $^1P$ | -2.05807 |
| -2.02045 | 0.58068 | 0.00456 | -0.81622 | 0.00119 | -2.05803 | $^1P$ | -2.05807 |

Table 3. Linear combinations, $\Phi$, of higher eigenfunctions with $\psi_0$, which are orthogonal to $\psi_1$, but have energy $E[\Phi] = E[\psi_1] = E_1$.

## Conclusions

The excited states $\psi_n$ energy are saddle points in the Hilbert space. Truncated functions orthogonal to some approximant $\phi_0$ of the ground state $\psi_0$ can lie either below or above the 1st excited state eigenfunction $\psi_1$ while the closest function to $\psi_1$, in the subspace that is orthogonal to $\phi_0$, lies below $\psi_1$. According to HUM theorem, any function obtained by optimization of the 2nd (1st excited) root of the secular equation has to lie above $\psi_1$. Hence, although, by construction, it is orthogonal to the 1st root, $\phi_0$, and although it is optimized, it is not the closest to $\psi_1$. Optimizing any root deteriorates all other roots, hence any optimized HUM excited state approximant is orthogonal to deteriorated lower roots. Thus, in order to approach the exact excited state eigenfunction (saddle point), it should rather be an as much as possible "large" expansion in the Hilbert space (the larger, the better): "Small" HUM truncated wave functions may be misleading. However, especially in large systems, "small" functions are useful. If one uses the proposed minimization functional $F_n$, (which - it is important to mention - *does not use any orthogonality to lower-lying approximate* wave functions, thus allowing approaching the exact Hamiltonian eigenfunction, even in small truncated spaces) then truncated functions $\phi_n$, that minimize $F_n$, approach the $\psi_n$ saddle point, irrespectively of whether they are "large" or "small". This was demonstrated in excited He $^1S$ and $^3S$ with both Hylleraas and Configuration Interaction truncated expansions. In minimizing $F_n$, at the minimum (near it within a tolerance criterion) that approaches the exact saddle point, some of our functions ("small" ones) stopped in the lower energy side of the saddle point. In such cases we used a reliability criterion: If ($F_n$ - $E[\phi_n]$) and an estimate of $L[\phi_n]$ (if one is available [cf. Eq. 2]) are small, then $\phi_n$ is reliable. Finally we demonstrated that infinitely many functions $\Phi$ orthogonal to the excited state $\psi_1$ can have the exact excited state energy $E = E[\psi_1]$ despite their orthogonality to $\psi_1$.

## ACKNOWLEDGMENTS

This work was sponsored by: Key Project of National Social Science (Grant No. 15AJL004), China, and by the General Secretariat for Research and Technology, Greece, through project "Advanced Materials and Devices" (MIS: 5002409).